\documentclass[aps,prb,reprint,superscriptaddress]{revtex4-1}

\usepackage{amsfonts} 
\usepackage{amsmath}
\usepackage{amssymb}
\usepackage{cancel}
\usepackage{bm}
\usepackage{color}
\usepackage[usenames,dvipsnames,svgnames,table]{xcolor}
\usepackage{graphicx}
\usepackage{amsfonts}
\usepackage{amsmath}
\usepackage[lofdepth,lotdepth,caption=false]{subfig}
\usepackage[varg]{txfonts}
\usepackage{wrapfig}
\usepackage[colorlinks = true,linkcolor = red,urlcolor=cyan,citecolor=red]{hyperref}

\begin{document}


\title{Magnetic excitation spectra in pyrochlore iridates}


\author{Eric Kin-Ho Lee}
\affiliation{Department of Physics and Centre for Quantum Materials,
  University of Toronto, Toronto, Ontario M5S 1A7, Canada}

\author{Subhro Bhattacharjee} \affiliation{Department of Physics and
  Centre for Quantum Materials, University of Toronto, Toronto,
  Ontario M5S 1A7, Canada} \affiliation{Department of Physics and
  Astronomy, McMaster University, Hamilton, Ontario L8S 4M1, Canada}

\author{Yong Baek Kim} \affiliation{Department of Physics and Centre
  for Quantum Materials, University of Toronto, Toronto, Ontario M5S
  1A7, Canada} \affiliation{School of Physics, Korea Institute for
  Advanced Study, Seoul 130-722, Korea}

\date{\today}

\begin{abstract}
  Metal-insulator transitions in pyrochlore iridates
  (A$_2$Ir$_2$O$_7$) are believed to occur due to subtle interplay of
  spin-orbit coupling, geometric frustration, and electron
  interactions. In particular, the nature of magnetic ordering of
  iridium ions in the insulating phase is crucial for understanding of
  several exotic phases recently proposed for these materials. We
  study the spectrum of magnetic excitations in the
  intermediate-coupling regime for the so-called all-in/all-out
  magnetic state in pyrochlore iridates with non-magnetic A-site ions
  (A=Eu,Y), which is found to be preferred in previous theoretical
  studies. We find that the effect of charge fluctuations on the
  spin-waves in this regime leads to strong departure from the
  lowest-order spin-wave calculations based on models obtained in
  strong-coupling calculations. We discuss the characteristic features
  of the magnetic excitation spectrum that can lead to conclusive
  identification of the magnetic order in future resonant inelastic
  x-ray (or neutron) scattering experiments. Knowledge of the nature
  of magnetic order and its low-energy features may also provide
  useful information on the accompanying metal-insulator transition.
\end{abstract}

\maketitle

\section{\label{sec:intro}Introduction}

Pyrochlore iridates (A$_2$Ir$_2$O$_7$) have recently attracted much
attention as prominent examples of $5d$ transition metal oxides where
interplay of spin-orbit (SO) coupling and electron interactions can
lead to a number of competing exotic
phases.\cite{2010_Nat_Machida,2007_PRL_Machida,
  2012_PRB_Witczak-Krempa_a, 2012_PRL_Go, 2010_PRB_Yang,
  2010_Nat_Pesin, 2011_PRB_Wan, 2008_PRL_Kim, 2009_Sci_Kim,
  2011_JPSJ_Matsuhira,2011_PRB_Zhao,
  2002_JPSJ_Fukazawa,2003_JPSJ_Aito, 2006_PRL_Nakatsuji,
  2011_PRB_Kargarian,2010_PRB_Yang,2008_PRB_Singh, 2011_JPSJ_Kurita,
  2012_PRB_Disseler_a, 2012_PRB_Disseler_b,
  2012_PRB_Tafti,2012_arX_Chen} Interestingly, most of these materials
are found to either exhibit a finite temperature metal-insulator (MI)
transition \cite{2001_JPSJ_Yanagishima, 2001_JPCM_Taira,
  2007_JPSJ_Matsuhira, 2011_JPSJ_Matsuhira, 2011_PRB_Zhao,
  2012_PRB_Ishikawa, 2012_JPSJ_Tomiyasu, 2012_PRB_Tafti}, or naturally
lie close to a zero temperature MI quantum phase transition (both
pressure-driven\cite{2012_PRB_Tafti, 2011_PRB_Sakata}, and/or
chemical-pressure-driven via variation of $A$-site
ions\cite{2001_JPSJ_Yanagishima, 2011_JPSJ_Matsuhira}).  It is now
believed that the nature of the MI transitions in these systems are
related to the magnetic order of iridium (Ir) ions in the
low-temperature insulating phase.\cite{2007_JPSJ_Matsuhira,
  2011_JPSJ_Matsuhira, 2012_PRB_Ishikawa, 2012_JPSJ_Tomiyasu} Further,
the details of such magnetic ordering pattern are known to be crucial
for some of the proposed novel phases like the Weyl
semimetal.\cite{2011_PRB_Wan, 2012_PRB_Witczak-Krempa_a} Therefore,
the determination of Ir magnetic configuration is important, both to
gauge the relevance of the proposed novel phases and to shed light on
the nature of the MI transition in pyrochlore iridates.  However,
conclusive experimental evidence for the nature of such magnetic order
in pyrochlore iridates with non-magnetic A-site ions (such as
Eu$_2$Ir$_2$O$_7$ and Y$_2$Ir$_2$O$_7$)\cite{2001_JPSJ_Yanagishima,
  2001_JPCM_Taira, 2002_JPSJ_Fukazawa, 2011_PRB_Zhao,
  2012_PRB_Ishikawa, 2012_PRB_Shapiro, 2012_PRB_Disseler_b} is
presently lacking.

\begin{figure}[h!]
\centering
\includegraphics[scale=0.8]{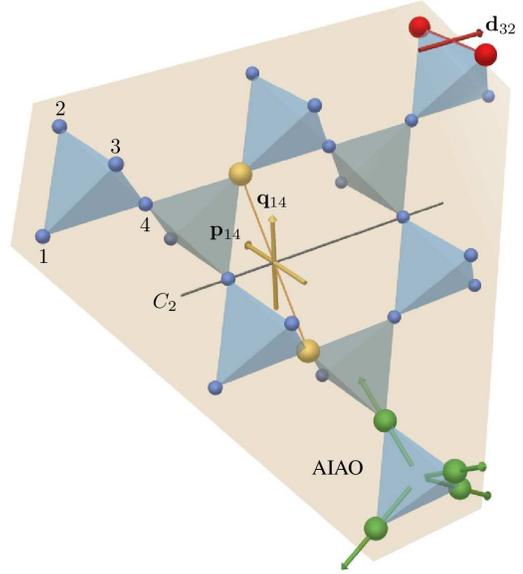}
\caption{\label{fig:pyrochlore}(Color online) The top left tetrahedron
  indicates the sublattice indices used in our work.  The red vector
  in the top right tetrahedron shows the nearest-neighbor
  $\mathbf{d}_{32}$.  The gold vectors $\mathbf{p}_{14}$ and
  $\mathbf{q}_{14}$ are perpendicular to the $C_2$ axis and span the
  plane in which next-nearest-neighbor $\mathbf{v}_{\langle\langle
    14\rangle\rangle}$ must lie.  The bottom right tetrahedron with
  the green vectors shows the all-in/all-out configuration.}
\end{figure}

Several important clues regarding the magnetic order have been
revealed by recent muon spin resonance/relaxation and magnetization
measurements on both Eu$_2$Ir$_2$O$_7$\cite{2011_PRB_Zhao,
  2012_PRB_Ishikawa} and Y$_2$Ir$_2$O$_7$\cite{2012_PRB_Disseler_b}:
in the low temperature insulating phase, these measurements suggest
that localized Ir moments exhibit long-range magnetic order that may
not break the pyrochlore lattice symmetry. These results are
consistent with the claim that in the ground state, the Ir moments
order in the non-collinear, all-in/all-out (AIAO) fashion (see
Fig. \ref{fig:pyrochlore}), as was previously found in calculations
for both the strong \cite{2005_PRB_Elhajal} and intermediate electron
correlation regimes.\cite{2011_PRB_Wan, 2012_PRB_Witczak-Krempa_a}
However, the above findings cannot conclusively prove that
Eu$_2$Ir$_2$O$_7$/Y$_2$Ir$_2$O$_7$ orders in the AIAO fashion; a study
of the low-energy magnetic excitations is required to identify the
signatures unique to the AIAO state.  Indeed, recent RIXS and
inelastic neutron scattering experiments have revealed the magnetic
excitation spectra of other iridates such as
Sr$_2$IrO$_4$\cite{2012_PRL_Kim_214},
Sr$_3$Ir$_2$O$_7$\cite{2012_PRL_Kim_327}, and
Na$_2$IrO$_3$\cite{2012_PRL_Choi}.


In this paper, we compute the magnetic excitation spectrum in both the
intermediate and strong electron correlation regimes.  Starting from
an effective Hubbard model in the $J_{\text{eff}}=1/2$ basis to
describe Ir electrons, we study the evolution of the magnetic
excitation spectrum in the intermediate-$U$ regime by computing the
transverse magnetic dynamic structure factor within the random phase
approximation (RPA).  The robust symmetry-protected degeneracies of
the spin-wave spectrum, the occurrence of Landau damping, as well as
the characteristic dispersion along high symmetry directions are
demonstrated to be the defining signatures of the AIAO state in the
intermediate-$U$ regime.  In the strong correlation limit, we derive
an effective spin-model and calculate the corresponding spin-wave
excitations. Through naive fitting of the RPA dynamic structure factor
with the strong-coupling spin-wave results, we find that the magnetic
excitation spectrum at intermediate-$U$ shows strong departures from
the lowest-order spin-wave calculations. Current experiments suggest
that the intermediate-coupling regime may be more suitable for
describing Eu$_2$Ir$_2$O$_7$, where the charge gap estimated from the
resistivity measurements is found to be small ($\sim
10$~meV).\cite{2011_PRB_Zhao}

The rest of this paper is organized as follows. We begin with a brief
description of the Hubbard model relevant to the pyrochlore iridates
in Section \ref{sec:model}.  We introduce a new parametrization for
the tight-binding parameters, which elucidates the structure of the
mean-field phase diagram presented in Section \ref{sec:mft}.
Subsequently, we discuss the results obtained both in the
intermediate-$U$ (Section \ref{subsec:intermediateu}) and large-$U$
(Section \ref{subsec:largeu}) regimes.  We also discuss the unique
signatures of the AIAO phase and compare the results obtained in the
two different regimes.  Implications of our results are summarized in
Section \ref{section_summarize}. Further details regarding various
calculations are given in the appendices.

\section{Microscopic Hamiltonian}
\label{sec:model}
In the pyrochlore iridates, $\text{Ir}^{4+}$ ions are located at the
center of corner-sharing oxygen octahedra.  This leads to crystal-field
splitting of the $5d$ orbitals into upper $e_g$ and lower $t_{2g}$
orbitals. The six-fold degenerate $t_{2g}$ orbitals (including spin
degeneracy) is then split into an upper $J_{\text{eff}}=1/2$ doublet
and lower $J_{\text{eff}}=3/2$ quadruplet by the atomic SO coupling
with energy separation of $3\lambda/2$ ($\lambda \approx 500$~meV for
Ir).  Therefore, in the atomic limit, the five valence electrons will
fully fill the $J_{\text{eff}}=3/2$ states, half-fill the
$J_{\text{eff}}=1/2$ states, and leave the $e_g$ orbitals
unoccupied. This atomic picture suggests that the low-energy physics
can be captured by considering only the half-filled
$J_{\text{eff}}=1/2$ states.\cite{2008_PRL_Kim, 2009_Sci_Kim}
Therefore it is useful to start from the most general on-site Hubbard
model in the $J_{\text{eff}}=1/2$ basis allowed by symmetry:
\begin{equation}
  \label{eq:hubbard_hamiltonian} 
  H=\sum_{ij}c_{i}^{\dagger}h_{ij}c_{j}+U\sum_{i}n_{i\uparrow}n_{i\downarrow},
\end{equation}
where $i,j$ are lattice site indices, $c_{i}=( c_{i\uparrow}, \:
c_{i\downarrow})$ are the electron annihilation operators, $\uparrow$
and $\downarrow$ are the z-components of the psuedospin operator
defined in the global basis, and
$n_{i\sigma}=c^\dagger_{i\sigma}c_{i\sigma}$ is the electron number
operator at site $i$ of psuedospin $\sigma$.  In general the hopping
matrix $h_{ij}$ is complex and can be constrained by considering
time-reversal invariance and various space group symmetries (Moriya
rules).\cite{1960_PR_Moriya, 2005_PRB_Elhajal, 2011_JPSJ_Kurita}

Time-reversal invariance restricts the hopping matrix to the form:
\begin{equation}
  h_{ij}=t_{ij} \mathbb{I} + i \mathbf{v}_{ij} \cdot \mathbf{\sigma},
\end{equation}
where $\mathbb{I}$ is the $2\times2$ identity matrix,
$\mathbf{\sigma}(=\sigma_x,\sigma_y,\sigma_z)$ are the Pauli matrices (in the pseudospin space),
and $t_{ij}$ and
$\mathbf{v}_{ij}=({\text{v}}^x_{ij},{\text{v}}^y_{ij},{\text{v}}^z_{ij})$
are real hopping amplitudes.  Hermiticity of the Hamiltonian implies
\begin{align}
t_{ij}=t_{ji}~~ {\text{and}}~~\mathbf{v}_{ij} =
-\mathbf{v}_{ji}.
\end{align}
$t_{ij}$ and $\mathbf{v}_{ij}$ transform as a scalar and as a
psuedovector respectively under the space group symmetries of the
lattice.\cite{1960_PR_Moriya, 2005_PRB_Elhajal} In particular these
symmetries constrain the nearest-neighbor (NN) $\mathbf{v}_{\langle ij
  \rangle}$s to be perpendicular to the mirror plane containing $i$
and $j$.  The two possible directions that $\mathbf{v}_{\langle ij
  \rangle}$ can point are classified as ``direct'' or
``indirect''.\cite{2005_PRB_Elhajal} We shall denote the unit vector
of the ``direct'' case as $\mathbf{d}_{ij}$ (see
Fig. \ref{fig:pyrochlore}). For next-nearest-neighbors (NNN)
$\langle\langle ij \rangle\rangle$, the two-fold rotation axis that
exchanges $i$ and $j$ restricts $\mathbf{v}_{\langle\langle ij
  \rangle\rangle}$ to lie in the plane normal to this
axis.\cite{2012_arX_Witczak-Krempa_b} We can parametrize this plane by
the two orthonormal vectors
\begin{align}
\mathbf{p}_{ij}\equiv\sqrt{6}/4\left(-\mathbf{R}_{ij}+\mathbf{D}_{ij}\right)\\
\mathbf{q}_{ij}\equiv\sqrt{3}/2\left(\mathbf{R}_{ij}+\mathbf{D}_{ij}\right),
\end{align}
where $\mathbf{R}_{ij}\equiv\mathbf{r}_{ik}\times \mathbf{r}_{kj}$ and
$\mathbf{D}_{ij}\equiv\mathbf{d}_{ik} \times \mathbf{d}_{kj}$,
site $k$ being the common NN of $i$ and $j$ (see
Fig. \ref{fig:pyrochlore}).

The above constraints allow us to parametrize $h_{ij}$ by two real NN
(unprimed) hopping amplitudes and three NNN (primed) hopping
amplitudes:
\begin{align}
  \label{eq:hoppings}
  h_{\langle ij\rangle}(t_1,t_2)&=t_1 \mathbb{I} + i t_2 \mathbf{d}_{ij} \cdot \mathbf{\sigma}, \nonumber \\
  h'_{\langle\langle ij\rangle\rangle}(t'_1,t'_2,t'_3)&=t'_1
  \mathbb{I} + i \left( t'_2 \mathbf{p}_{ij} + t'_3
    \mathbf{q}_{ij}\right) \cdot \mathbf{\sigma}.
\end{align}

For the rest of the paper, instead of using
$(t_1,t_2),(t'_1,t'_2,t'_3)$ as our hopping parameters, we find it
more convenient to work with $(t, \theta), (t', \theta', \phi')$ which
makes certain symmetries of the phase diagram readily
accessible. These are defined as
\begin{equation}
  \label{eq:nnhopping}
  \left(
    \begin{matrix}
      t_1 \\
      t_2
    \end{matrix}
  \right)
  =
  t\left(
    \begin{matrix}
      \cos(\theta_t/2-\theta) \\
      \sin(\theta_t/2-\theta)
    \end{matrix}
  \right),
\end{equation}
\begin{equation}
  \label{eq:nnnhopping}
  \left(
    \begin{matrix}
      t'_1 \\
      t'_2 \\
      t'_3
    \end{matrix}
  \right)
  =
  t'\left(
    \begin{matrix}
      \cos(\theta_t/2-\theta')\sin(\phi')\\
      \sin(\theta_t/2-\theta')\sin(\phi')\\
      \cos(\phi')
    \end{matrix}
  \right),
\end{equation}
where $\theta_t=2\arctan(\sqrt{2})\approx109.47^{\circ}$ is the
tetrahedral angle.  This parametrization together with the definition
of $\mathbf{p}_{ij}$ and $\mathbf{q}_{ij}$ naturally makes the
following property of the model manifest: if we perform the basis
transformation $c_{ia\alpha}\rightarrow e^{-i\pi \hat{n}_{a}\cdot
  \mathbf{\sigma}_{\alpha\beta}}c_{ia\beta}$, where $\hat{n}_a$ is the
unit vector pointing from sublattice $a$ to the center of the
tetrahedron, the hopping parameters $\theta$ and $\theta'$ are
transformed to $-\theta$ and $-\theta'$ respectively.  In other words,
the Hamiltonian parametrized by hopping parameters $(t, \theta), (t',
\theta', \phi')$ yields identical features as the Hamiltonian
parametrized by $(t, -\theta),(t',-\theta',\phi')$.

We can also derive hopping matrices obeying the above symmetry
constraints by considering the Slater-Koster approximation of orbital
overlaps.  In terms of this microscopic approach, the relevant
Slater-Koster parameters are: (1) Ir-Ir hopping via direct overlap of
$d$-orbitals. There are three such overlaps: $t_{\sigma}$ (for
$\sigma$-bonds), $t_{\pi}$ (for $\pi$-bonds), and $t_\delta$ (for
$\delta$-bonds) for NN (and primed ones for NNN), and (2) $t_o$
describing the hopping between the Ir atoms via the intermediate
oxygen. Details of such a derivation can be found in
Ref. \onlinecite{2010_Nat_Pesin,
  2012_PRB_Witczak-Krempa_a,2012_arX_Witczak-Krempa_b} and Appendix
\ref{sec:slater-koster}.
\section{\label{sec:mft}Mean-Field Phase Diagram}

\begin{figure}
\includegraphics[width=3.4in]{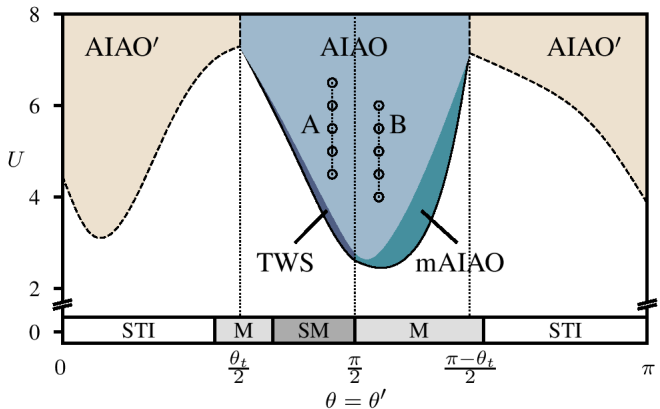}
\caption{\label{fig:pd}(Color online) Mean-field phase diagram with
  hopping amplitudes $\theta'=\theta$,
  $t'/t=0.1$, $t=1$, and $\phi'=5\pi/6$.  The phases in the non-interacting
  limit is shown at $U=0$. First order transitions are indicated by
  dashed lines while second order transitions are indicated by solid
  black lines.  Lines A ($\theta=1.45$) and B ($\theta=1.70$) represent the two
  cuts for which the RPA dynamic structure factors are calculated (Fig. \ref{fig:sf}).}
\end{figure}

In Fig. \ref{fig:pd}, we show the mean-field phase diagram for
$0<\theta<\pi$.  We have chosen $\theta=\theta'$, $t'/t=0.1$ and
$\phi'=5\pi/6$. Throughout our calculation we have set $t=1$ as our
energy scale and we comment on its possible values in physical systems
in the concluding section. We note that this parametrization is
somewhat different from those used in
Ref. \onlinecite{2012_PRB_Witczak-Krempa_a} where the Ir-O-Ir hopping
amplitude was set as the unit of energy. The phase diagram for
$0>\theta>-\pi$ will be identical due to the $(t, \theta), (t',
\theta', \phi')\sim\!(t, -\theta),(t',-\theta',\phi')$ structure as
discussed in the Section \ref{sec:model}. Our phase diagram is
consistent with those obtained in previous
calculations.\cite{2012_PRB_Witczak-Krempa_a}

In the non-interacting limit, depending on the value of $\theta$, we
find a strong topological-insulator (STI), the metallic (M), and the
semimetallic phase (SM).  The M phase has small particle- and
hole-like pockets, whereas the SM phase has a quadratic-band-touching
at the $\Gamma$-point at the chemical potential.  At finite $U$, only
two magnetic configurations are found: the all-in/all-out (AIAO) and
the rotated all-in/all-out (AIAO$'$)
orders.\cite{2012_PRB_Witczak-Krempa_a} The AIAO configuration is
realized by increasing $U$ starting with the SM or M phase.  On the
other hand, the AIAO$'$ configuration is realized by increasing $U$ in
the M or STI phase.  Phase transitions to the AIAO$'$ by an increase
in $U$ is of first order.  Also, the transitions between the AIAO and
the AIAO$'$ phases are of first order and occur at $\theta_t/2$ and
$\pi-\theta_t/2$.  In the non-interacting limit, band-inversion at the
$\Gamma$-point also occurs at these values of $\theta$.  All other
transitions are of second order.

At large $U$'s, the system is gapped, while for $U$ values near the
onset of magnetic order, the single-particle spectrum may continue to
remain gapless even after the onset of magnetic order.  The
topological Weyl semimetal (TWS) and the magnetically-ordered metallic
(mAIAO) phases are realized in this gapless window---the former is
developed via the splitting of the quadratic-band touching at the
Fermi-level of the SM phase while the latter is realized due to the
presence of particle-hole pockets in the M phase.
\section{\label{sec:spinwaves} Magnetic Excitation spectrum}

As pointed out in Sec. \ref{sec:mft}, for a sufficiently large Hubbard
interaction $U$ and for $\theta_t/2<\theta<\pi-\theta_t/2$, the
mean-field solution is the AIAO state.  We want to study the nature of
the low-energy magnetic excitation spectrum of this
magnetically-ordered state, which are composed of the transverse
fluctuations of the spins about their local ordering directions. We
study these excitations in both the intermediate and strong electron
correlation regimes and we accomplish this by applying two different
and contrasting approaches.  For the case of intermediate-$U$, we
study the spin-waves by computing the RPA transverse spin-spin dynamic
structure factor at zero temperature.  In the large-$U$ regime, we
perform a strong-coupling expansion of our Hubbard model
(Eq. \ref{eq:hubbard_hamiltonian}) to derive an effective spin-model.
The spin-model is then used to calculate the spin-wave spectrum about
the AIAO state within the Holstein-Primakoff approximation.
\subsection{\label{subsec:intermediateu}Intermediate-$\mathbf{U}$: RPA dynamic
  structure factor}
The information about the spin-waves are contained in the transverse
part of the RPA dynamic spin-spin susceptibility matrix. This is given
by:
\begin{equation}
  \label{eq:RPA}
  \chi_{\text{RPA}\perp}(\mathbf{q},\omega)
  =(\mathbf{1} - U \chi_{\text{MF}\perp}(\mathbf{q},\omega))^{-1}(\chi_{\text{MF}\perp}(\mathbf{q},\omega)),
\end{equation}
where $\chi_{\text{MF}\perp}$ is the bare mean-field transverse
spin-spin susceptibility, and $U$ is the Hubbard repulsion.  Since the
pyrochlore unit cell has four sublattices, $\chi_{\text{MF}\perp}$ is
a $4\times4$ matrix.  We compute the trace of the imaginary part of
the RPA susceptibility, \emph{i.e.}, the RPA dynamic structure
factor. This trace sums over the contribution of the individual
spin-wave bands (there are four such bands) and gives the overall
intensity that will be observed in inelastic neutron scattering or
RIXS experiments. The details of the form of the susceptibility matrix
is discussed in Appendix \ref{sec:mf_sus}.

\begin{figure*}
  \centering
  \subfloat[][]{
    \includegraphics[width=3.4in,trim=10pt 0 2pt 0]{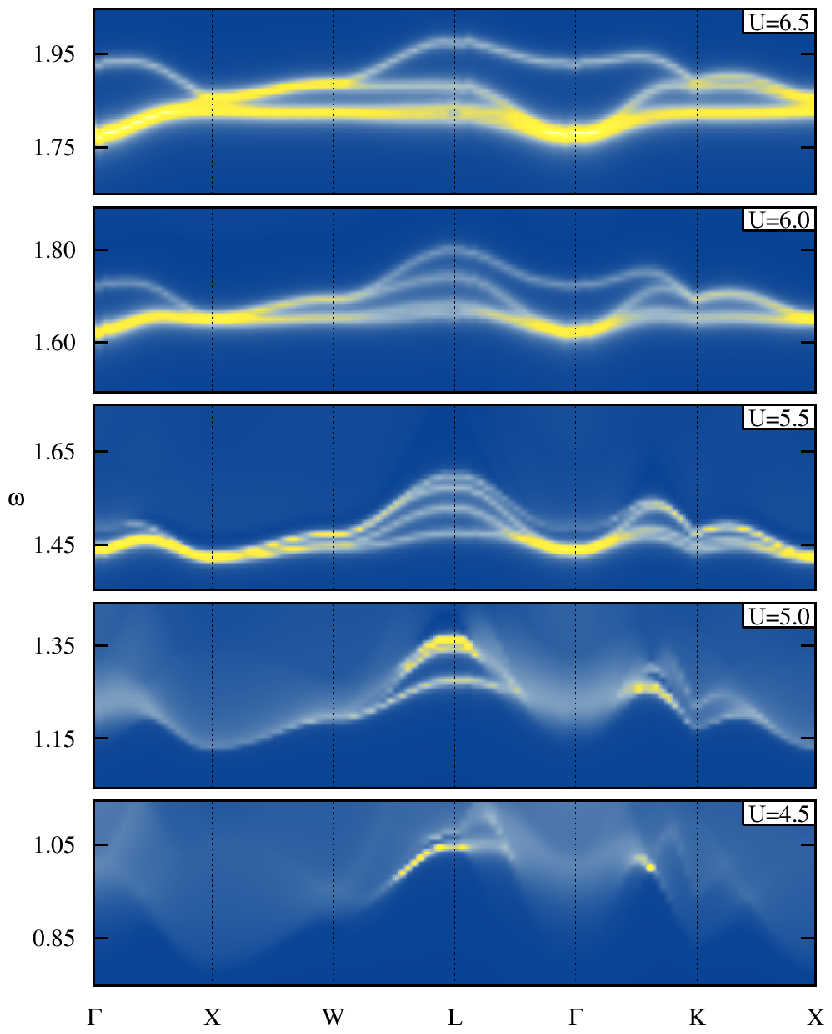}
    \label{fig:q}
  }\;\:
  \subfloat[][]{
    \includegraphics[width=3.4in,trim=10pt 0 2pt 0]{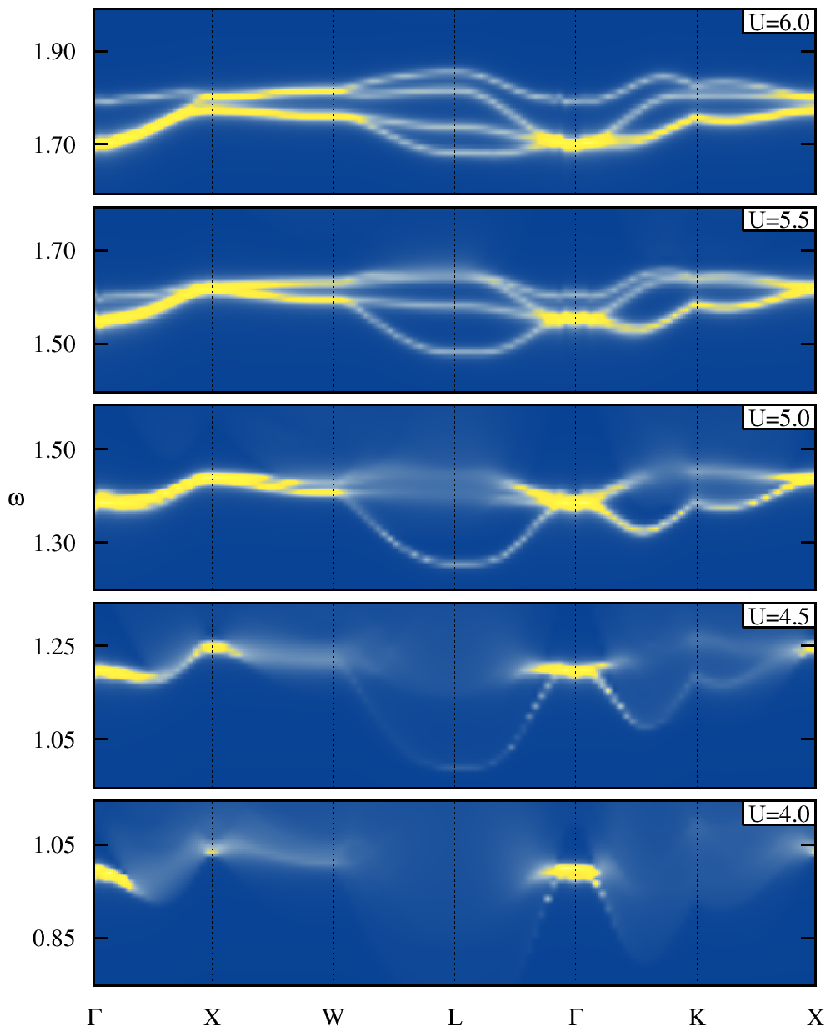}
    \label{fig:m}
  }
  \caption{(Color online) RPA dynamic structure factors for
    (a) $\theta=1.45$ and (b) $\theta=1.70$ for
    various values of $U$; $\omega$ is in units of $t=1$.  These
    correspond to cuts A and B in Fig. \ref{fig:pd} respectively.
    Sharp dispersions can be seen at larger values of $U$'s, while
    lighter intensity and broadened spectra is seen due to Landau
    damping in parts of the Brillouin zone at lower $U$'s.
    Degeneracies at $\Gamma$, X, and W are symmetry-protected and are
    robust features of the AIAO state.}
  \label{fig:sf}
\end{figure*}
\paragraph*{Results:} We consider the RPA dynamic
structure factor along two representative cuts (A and B) in the phase
diagram as $U$ is increased.  For cut A, $\theta=1.45$ while for cut
B, $\theta=1.70$ as shown in Fig. \ref{fig:pd}.  Due to the presence
of spin-orbit coupling ($t_2\neq 0$), spin-rotation symmetry is
explicitly broken, and the spin-wave spectra are expected to be
gapped.

The dynamic structure factors for cut A are shown in Fig. \ref{fig:q}.
For this cut, the system is in the quadratic-band-touching SM phase in
the non-interacting limit.  For $U=4.5$, the low-lying particle-hole
continuum damps and broadens the spin-wave spectrum throughout most of
the Brillouin zone.  This so-called Landau damping occurs because
spin-waves decay through interactions with single-particle excitations
and acquire a finite lifetime, which broadens its
spectrum\cite{2011_PRB_Buczek,2003_Mohn}. The damping does not
occur near the $L$-point, where the dispersions extend out of the
continuum and produce sharp bands.  As $U$ is increased, the
particle-hole continuum is shifted upwards in energy throughout the
Brillouin zone, revealing all four spin-wave modes.  The lower energy
modes are relatively dispersionless compared to the higher energy
modes, which disperses most markedly near the $L$-point.  At $U=6.5$,
the degeneracies of the spectrum become more apparent: there are two
two-fold degeneracies at the $X$- and $W$-points, while the
$\Gamma$-point has a three-fold degeneracy.  We note that for
$U\lesssim 4.5$, which includes the TWS phase, all the spin-wave modes
are damped by the particle-hole continuum.

For the second cut (Fig. \ref{fig:m}), the system is in the metallic
phase in the non-interacting limit.  Well-defined spin-wave
excitations are only observed for $U\gtrsim 4.0$. Starting at $U=4.0$,
the spin-wave modes appear near the $\Gamma$ point.  At $U=4.5$,
low-energy, damped features can be seen near the $L$-point and along
the $\Gamma-K$ line.  For $U=5.0$, most of the spin-wave modes become
sharply defined as the particle-continuum shifts upward.  As $U$ is
increased further, the low-lying dispersion at $L$ shifts up and bands
at $\Gamma$ begin to separate while maintaining the three-fold
degeneracy required by symmetry.  At $U=6.0$, the spectrum begins to
resemble the $U=6.5$ spectrum of the first cut.

Degeneracies at the high symmetry points $\Gamma$, X, and W are
symmetry-protected, and therefore, they are characteristic to the AIAO
state, which preserves the lattice symmetry.  The almost flat
dispersion encountered at the zone boundary (X-W) is also a
distinguishing feature.
\subsection{\label{subsec:largeu}Strong-Coupling Expansion - Linear
  Spin-Wave Theory}
We now look at the spin-wave spectrum in the strong-coupling limit of
large $U/t$.  In this limit and at half-filling, we can apply
perturbation theory to obtain the following effective spin Hamiltonian
at the lowest order.\cite{2010_Nat_Pesin, 2012_PRB_Witczak-Krempa_a}
\begin{align}
  \label{eq:spin_hamiltonian}
  H_{\text{spin}}&=\sum_{ij}\Lambda^{ab}_{ij}S^a_{i}S^b_{j} \nonumber \\
  &=\sum_{ij}\left(J \mathbf{S}_{i} \cdot \mathbf{S}_{j}
    + \mathbf{D}_{ij} \cdot \mathbf{S}_{i} \times \mathbf{S}_{j}
    + S^{a}_{i}\Gamma^{ab}_{ij}S^{b}_{j}\right),
\end{align}
where the three terms in the last line are the trace, traceless
antisymmetric, and traceless symmetric parts of $\Lambda^{ab}_{ij}$.
These terms correspond to the Heisenberg, the Dzyaloshinski-Moriya
(DM), and the anisotropic interactions, respectively, and are related
to the hopping amplitudes of Eq. (\ref{eq:hubbard_hamiltonian}) by:
\begin{align}
  \label{eq:jdgamma}
  J&= 4U^{-1}\left(t^2-|\mathbf{v}|^2/3\right), \nonumber \\
  \mathbf{D}_{ij}&=8U^{-1}t\mathbf{v}_{ij}, \nonumber \\
  \Gamma^{ab}_{ij}&=8U^{-1}\left({\text{v}_{ij}}^{a}{\text{v}_{ij}}^{b}-\delta^{ab}|\mathbf{v}|^2/3\right),
\end{align}
(the magnitude of ${\bf v}_{ij}(=|{\bf v}|)$ is site independent)
which holds for both NN and NNN hopping amplitudes.  The ``direct"
configuration of the DM vectors are known to stabilize the AIAO
state\cite{2005_PRB_Elhajal} which is in agreement with our earlier
mean-field phase diagram. Hence we consider the low-energy spin-wave
expansion about the AIAO state for the above spin Hamiltonian.

\begin{figure}
\includegraphics{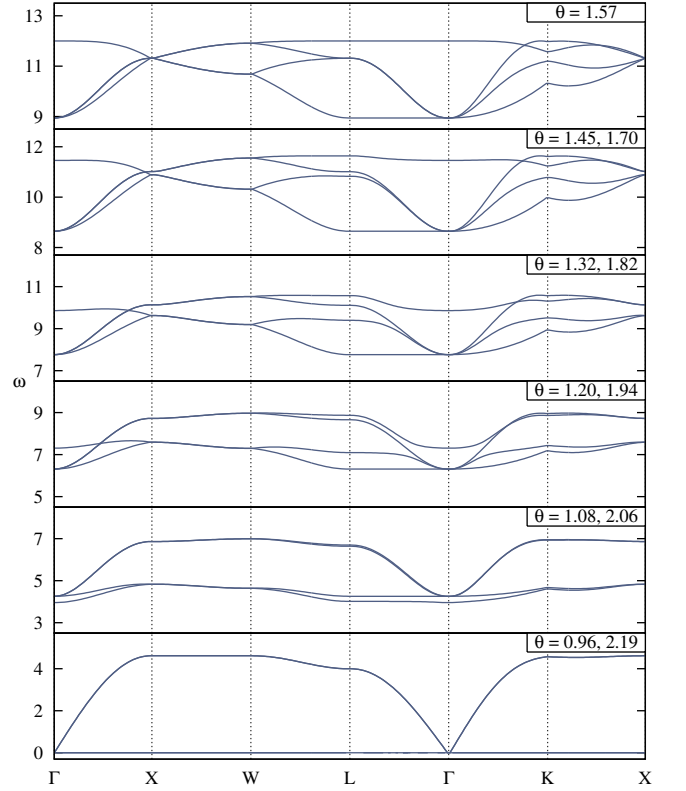}
\caption{\label{fig:hp_spectra}Evolution of the Holstein-Primakoff
  spin-wave spectra as angle $\theta$ is varied between
  $(\theta_t/2\approx 0.96)<\theta<(\pi-\theta_t/2\approx 2.19)$;
  $\omega$ is in units of $t^2/U$.  $\theta$ and $\pi-\theta$ yield
  identical spin-wave spectra and this redundancy is reflected by each
  plot having two $\theta$ values.  Like the RPA results in
  Fig. \ref{fig:sf}, degeneracies at $\Gamma$, X, and W are
  symmetry-protected.}
\end{figure}

To obtain the spin-wave expansion about the AIAO state which orders
non-collinearly, we rotate our spin quantization axis locally in
alignment with the magnetic ordering.\cite{2011_PRX_Ross} To this end,
we define rotated spin operators, $\tilde{\mathbf{S}}$, such that
their local $S_z$ points to the direction of magnetic ordering at that
site.
\begin{equation}
  S^{a}_{i}=\left(\mathbf{R}_{i}(\tilde{\mathbf{S}}_{i})\right)^{a}=R^{ab}_{i}\tilde{S}^{b}_{i},
\end{equation}
where $\mathbf{R}_{i}$ and $R^{ab}_{i}$ are the rotation operator and
its matrix representation that takes the direction of magnetic order at site
$i$ and rotates it to the z-axis of the global coordinate system. With
these rotated operators, we can rewrite
Eq. (\ref{eq:spin_hamiltonian}):
\begin{equation}
  H_{\text{spin}}=\sum_{ij}\left[R^{\text{T}}_{i}\Lambda_{ij} R_{j}\right]^{ab}\tilde{S}^a_{i}\tilde{S}^b_{j},
\end{equation}
where $R^{\text{T}}$ indicates matrix transposition. 

After recasting our spin operators in the rotated coordinate system,
we are in the position to analyze the spin-waves about the
AIAO state by applying linear spin-wave theory.  First, we
rewrite our spin operators in the Holstein-Primakoff bosonic
representation:
\begin{align}
  \tilde{S}^{+}_{i}&=\sqrt{2s-a^{\dagger}_{i}a_{i}}a_{i}, \\ \nonumber
  \tilde{S}^{-}_{i}&=a^{\dagger}_{i}\sqrt{2s-a^{\dagger}_{i}a_{i}}, \\ \nonumber
  \tilde{S}^{\text{z}}_{i}&=s-a^{\dagger}_{i}a_{i},
\end{align}
where $s$ is the total spin angular momentum and we have introduced
four flavors of bosons, one for each sublattice of the pyrochlore
unit cell.  Next, we expand and truncate the spin Hamiltonian to
quadratic order, Fourier transform the bosonic operators, and solve
for the resulting excitation spectrum via a Bogoliubov transformation.

\paragraph*{Results:}We consider the spin-wave spectra obtained from
the effective spin Hamiltonian generated by only NN hopping
amplitudes. Adding up to $t'=0.1~ t$ NNN hopping amplitudes (not
shown) only leads to small changes (see below). In
Fig. \ref{fig:hp_spectra}, we depict the evolution of the spin-wave
spectrum as the NN hopping parameter $\theta$ is varied.  In the
absence of NNN hopping amplitudes, $\theta$ and $\pi-\theta$ yield
identical spin-wave spectra.  This structure is again related to our
choice of angular parametrization: $\theta$ and $-\theta$ are related
by a basis transformation and should therefore have the same spin-wave
spectrum.  Moreover, $-\theta\rightarrow-\theta+\pi$ is equivalent to
$t\rightarrow-t$ and $\mathbf{v}\rightarrow-\mathbf{v}$, which leaves
$J$, $\mathbf{D}_{ij}$, and $\Gamma_{ij}$ invariant.  Hence, $\theta$
and $\pi-\theta$ yield the same spin-wave spectrum and this fact is
noted by assigning each of the plots in Fig. \ref{fig:hp_spectra} with
two $\theta$ values.

For $\theta_t/2<\theta<\pi-\theta_t/2$, the spin-wave spectrum is
gapped and the gap decreases as we approach either endpoints of the
interval.  At the endpoints, two of the four bands of the spectrum
become both gapless and dispersionless, while outside the endpoints,
the lowest bands become negative in energy, signaling an instability of the AIAO state.
The onset of these instabilities is consistent with NN mean-field
theory results, which predicts first order transitions (between the
AIAO to the AIAO$'$ phase) at $\theta=\theta_t/2$ and $\theta =
\pi-\theta_t/2$.

We note that the degeneracies at the high-symmetry points $\Gamma$, X,
and W are consistent with the RPA results as they are protected by
symmetry.  Also, the flat dispersions at the zone boundary (X-W) is
also encountered in the present spin-wave calculation.  On the other
hand, the lowest energy dispersion along the L-$\Gamma$ line is
absolutely flat in this NN model. However, on adding small NNN hopping
(up to $t'=0.1~ t$), it acquires small dispersion. This should be
contrasted with RPA results in Fig.  \ref{fig:sf} where modes along
the L-$\Gamma$ line are more strongly dispersive.
\subsection{Comparison of RPA and strong-coupling results}

As the ratio of typical hopping scale to the Hubbard repulsion scale
($t/U$) increases, the higher-order contributions to the
strong-coupling perturbative expansion (in $t/U$) become increasingly
important and the strictly NN model we employed in Section
\ref{subsec:largeu} becomes inadequate in describing the magnetic
excitations of the AIAO state.  Not only do higher-order contributions
generate further-neighbor Heisenberg exchanges, ring-exchange type
terms arise and lead to renormalization of NN quadratic terms at the
linear spin-wave level.\cite{2004_PRB_Chernyshev, 2005_PRB_Delannoy}
Therefore, we should not expect a perfect agreement between the RPA
and strong-coupling results.

Nevertheless, we attempt to fit (by eye estimation) the RPA results
for large $U$ ({\it e.g.}, $U=6.0$ in cut B) with a linear spin-wave
(LSW) spectrum.  First, we fit the RPA dispersion features and overall
bandwidth, resulting in $J=0.13$, $|\mathbf{D}|=0.13$, $|\Gamma|=0.07$
(see Eq. \ref{eq:jdgamma} for definitions).  These parameter values
are different from those obtained from Eq. \ref{eq:jdgamma} which is
based strictly on the leading order strong-coupling expansion for the
NNs.  The gap obtained from the RPA calculation is
$\Delta_{\text{RPA}}\approx1.70$.

The resulting LSW fit captures the dispersion along $\Gamma$-X-W quite
well, but fails to capture the low-lying modes along the L-$\Gamma$
line and, in general, the higher energy modes where the fit at best is
qualitative.  We would also like to point out that fits at lower $U$
and along cut A have been attempted but large discrepancies in both
the features of the spectrum and the spin-wave gap have been found.

The above fitting results points out that the NN spin-model (and also
the NNN spin-model with up to $10\%$ NNN hopping amplitude) is grossly
inadequate to fit the RPA spin-wave spectrum quantitatively for
parameter values that encompasses the regime appropriate for the
pyrochlore iridates. In this context we would like to point out that
recent estimates of small charge gap ($\sim\!10$~meV) from the
resistivity measurements in Eu$_2$Ir$_2$O$_7$
\cite{2011_PRB_Zhao,2012_PRB_Ishikawa} seems to suggest that the
intermediate-coupling calculations may be better suited to describe
this compound.

\begin{figure}
  \includegraphics[width=3.4in,trim=5pt 0 2pt 0]{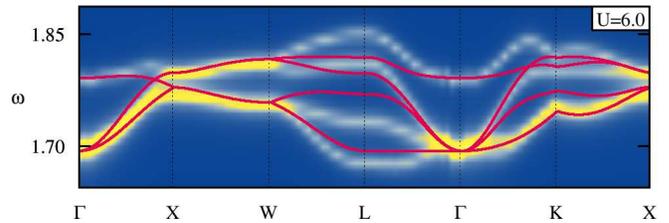}
  \caption{\label{fig:fit}(Color online) RPA results at $U=6.0$ for cut B (blue and
    yellow color-map) overlaid with a fit with the NN linear spin-wave
    spectrum (red); $\omega$ is in units of $t=1$.  Though some
    features along the $\Gamma$-X-W line can be fitted, other
    high-symmetry lines show a larger discrepancy.}
\end{figure}

\section{Summary}
\label{section_summarize}
To summarize, we have calculated the structure of the magnetic
excitation spectrum for the AIAO state that has been proposed for the
pyrochlore iridates such as Eu$_2$Ir$_2$O$_7$ or Y$_2$Ir$_2$O$_7$. For
intermediate correlations, we have calculated the transverse spin-spin
dynamic structure factor within the RPA approximation.  Features
particular to the AIAO configuration that can lead to conclusive
identification of the magnetic order in Eu$_2$Ir$_2$O$_7$ and
Y$_2$Ir$_2$O$_7$ were discussed.  For the large-$U$ limit, we used a
strong-coupling expansion to derive a spin-model and calculated the
linear spin-wave spectrum using the Holstein-Primakoff approximation.
By fitting the RPA results with the linear spin-wave spectrum, we
showed that results in the intermediate correlation regime are
substantially different from those obtained from the strong-coupling
theory.

Finally, from our calculations, we can make rough estimates of the
experimental energy scales for the excitation gap and dispersion
bandwidth by using Slater-Koster parametrization of orbital overlaps
(the relation connecting the symmetry-allowed parameters in
Eq. \ref{eq:hoppings} to Slater-Koster parameters are detailed in
Appendix \ref{sec:slater-koster}).  Similar to
Refs. \onlinecite{2012_PRB_Witczak-Krempa_a,
  2012_arX_Witczak-Krempa_b}, we have used $t_{\pi}=-2t_{\sigma}/3$,
$t_{\delta}=0$, while $t_\sigma=-1.16 t_o (-1.37 t_o)$ for cut $A(B)$
(the different orbital overlaps were introduced towards the end of
Section \ref{sec:model} and are also used in Appendix
\ref{sec:slater-koster}). Typically, the value of $t_o$---the Ir-O-Ir
hopping---is about $200$--$350$~meV in iridates with octahedral oxygen
environments.\cite{2010_PRB_Norman, 2012_PRL_Kim} Here we choose a
representative value of $t_o\approx300$~meV.  With these values, in
Fig. \ref{fig:sf}, the spin gap found from our RPA calculation is on
the order of $100$~meV and the dispersion width is on the order of
$\sim\!15$~meV. This value of spin gap is found to be very sensitive
to the value of $U$ while the bandwidth always remains in the same
regime. While a $100$~meV gap can be resolved within current RIXS
resolution, the dispersion width may be presently on the borderline of
resolvability.

In other iridium compounds, recent progress have been made in
RIXS\cite{2012_PRL_Kim_214, 2012_PRL_Kim_327, 2012_arX_Gretarsson},
neutron scattering\cite{2012_PRL_Choi, 2012_PRB_Ye}, resonant magnetic
x-ray scattering (RMXS)\cite{2011_PRB_Liu, 2012_arX_Clancy_b}, and
x-ray absorption spectroscopy (XAS)\cite{2012_arX_Clancy_a}
experiments.  In particular, RIXS and inelastic neutron scattering
experiments have recently revealed the magnetic excitation spectra of
Sr$_2$IrO$_4$\cite{2012_PRL_Kim_214},
Sr$_3$Ir$_2$O$_7$\cite{2012_PRL_Kim_327}, and
Na$_2$IrO$_3$\cite{2012_PRL_Choi}.  Future applications of these
techniques and improvements in experimental resolution may help reveal
the magnetic behavior and, in particular, the magnetic excitation
spectra of pyrochlore iridates, thereby conclusively determining the
nature of their magnetic order.

\begin{acknowledgements}
  We thank W. Witczak-Krempa, A. Go, Y.-J. Kim, P. Clancy, and
  B.J. Kim for useful discussion. This research was supported by the
  NSERC, CIFAR, and Centre for Quantum Materials at the University of
  Toronto.  Computations were performed on the gpc supercomputer at
  the SciNet HPC Consortium.\cite{2010_JPCS_Loken} SciNet is funded
  by: the Canada Foundation for Innovation under the auspices of
  Compute Canada; the Government of Ontario; Ontario Research Fund -
  Research Excellence; and the University of Toronto.
\end{acknowledgements}
\appendix

\section{\label{sec:slater-koster}Microscopic origins of hopping parameters}

In order to estimate the values of the hopping parameters, we turn to
a microscopic analysis of hopping paths.  We briefly discuss the
results of such an analysis and refer to
Ref. \onlinecite{2010_PRB_Yang, 2010_Nat_Pesin,
  2012_arX_Witczak-Krempa_b} for more details.

We consider two types of hopping: Ir-Ir hopping via overlap of
$d$-orbitals and O-Ir hopping between $p$-orbitals of O and
$d$-orbitals of Ir.  We parametrize $d$-$d$ overlaps with
Slater-Koster amplitudes $t_{\sigma}$, $t_{\pi}$, and $t_\delta$ (and
primed ones for NNN), while for $p$-$d$ overlaps, we parametrize them
with amplitudes $t_{pd\sigma}$, $t_{pd\pi}$, and O-Ir occupation
energy difference $\epsilon$. Here the subscripts $\sigma,\pi$ and
$\delta$ denotes the type of overlap of the orbitals. Also, in this
microscopic picture, we always work in the local axes defined by the
oxygen octahedra surrounding each Ir.

To arrive at a $J_{\text{eff}}=1/2$ model, we first employ
second order perturbation on the O-Ir hopping to generate an effective
NN Ir-Ir hopping between $d$-orbitals. This {\it indirect} hopping is given by
\begin{align}
t_{\text{o}}=t^2_{pd\pi}/|\epsilon|
\end{align} 
where $\epsilon$ is the difference of the on-site charging energies
between the oxygen $2$p orbitals and Ir $5$d orbitals. (This indirect
hopping receives contribution from the $t_{pd\sigma}$ overlap in
presence of distortion in the oxygen octahedra). We now project the
$d$-orbitals into the local $t_{2g}$ and, finally, into the local
$J_{\text{eff}}=1/2$ basis, to find the hopping matrix which has the
form given by Eq. (\ref{eq:hoppings}), where the relation with the
effective hopping parameters and more microscopic Slater-Koster
parameters is given by the following relations
\begin{align}
  \label{eq:param_convert}
  t_1&=\frac{1}{972}\left(51 t_{\sigma}-316 t_{\pi}-43t_{\delta}+520t_{\text{o}}\right), \nonumber \\
  t_2&=\frac{\sqrt{2}}{972}\left(60t_{\sigma}-160t_{\pi}-220t_{\delta}+112t_{\text{o}}\right), \nonumber \\
  t'_1&=\frac{1}{8748}\left(699t'_{\sigma} - 1628t'_{\pi} - 1843 t'_{\delta}\right), \nonumber \\
  t'_2&=\frac{\sqrt{2}}{8748}\left(-156t'_{\sigma} - 2720 t'_{\pi} - 4 t'_{\delta}\right),\nonumber \\
  t'_3&=\frac{1}{8748}\left(-144t'_{\sigma} - 960 t'_{\pi} + 1104 t'_{\delta}\right),
\end{align}
  From these
relations, it is straightforward to use Eqs. \ref{eq:nnhopping} and
\ref{eq:nnnhopping} to relate the angular parameters to the above
Slater-Koster parameters.

The range of physical NN hopping parameters explored in
Ref. \onlinecite{2012_arX_Witczak-Krempa_b} ($-1.2\lesssim
t_\sigma\lesssim-0.5$, $t_\pi=-2t_\sigma/3$, $t_\delta=0$ and $t_o=1$)
corresponds to $0.85 \lesssim\theta\lesssim 1.51$ with the appropriate
energy scaling of $t$.

\section{\label{sec:mf_sus}Mean-field transverse spin-spin
  susceptibility}
As Eq. \ref{eq:RPA} indicates, we need to compute the mean-field
transverse spin-spin susceptibility in order to obtain the RPA
susceptibility.  Here we provide details of such calculation for
non-collinear magnetic order like the AIAO state.

The mean-field susceptibility matrix is given by:
\begin{equation}
  \mathcal{\chi}^{ab}_{\text{MF}\perp}(\mathbf{q},t)=-i\Theta(t)\left\langle \left[ (\mathbf{S}^{a}_{\perp}(\mathbf{q},t))_{i}, (\mathbf{S}^{b}_{\perp}(-\mathbf{q},0))_{i} \right] \right\rangle ,
\end{equation}
where $\mathbf{S}^a_{\perp}(\mathbf{q},t)$ denotes the Fourier
transform of the component of the spin operator that is perpendicular
to the magnetic ordering direction for sublattice $a$ (see
Fig. \ref{fig:pyrochlore} for sublattice convention), and $i$ indexes
the components of $\mathbf{S}^a_{\perp}(\mathbf{q},t)$, which are to
be summed over.  If the direction of magnetic moment on sublattice is
given by the unit vector $\hat{n}^a$, then the transverse spin
operator for that sublattice is given by
\begin{equation}
  \mathbf{S}^a_{\perp}(\mathbf{q}) = -\hat{n}^a \times (\hat{n}^a \times \mathbf{S}^a(\mathbf{q})),
\end{equation}
where $\mathbf{S}^a(\mathbf{q})$ is the Fourier transform of the spin
operator. For the AIAO state, $\hat{n}^a$ points along the local [111]
direction of the lattice.

We write the spin operator with electron operators as
\begin{align}
\mathbf{S}^a(\mathbf{q})=\frac{1}{2}\sum_{\bf k} c^\dagger_{{\bf k}+{\bf q},\alpha}\boldsymbol{\sigma}_{\alpha\beta}c_{{\bf k},\beta}
\end{align}
where $\alpha,\beta=\uparrow,\downarrow$. Using this we get
\begin{align}
  &\mathcal{\chi}^{ab}_{\text{MF}\perp}(\mathbf{q},t) = -i M^{ab}_{\alpha\beta\gamma\delta}
  \Theta(t) \\
  &\times \sum_{\mathbf{k}_1 \mathbf{k}_2} \left\langle \left[ c_{a\alpha}^{\dagger}(\mathbf{q}+\mathbf{k}_1,t)
    c_{a\beta}(q_1,t), c_{b\gamma}^{\dagger}(-\mathbf{q}+\mathbf{k}_2,0) c_{b\delta}(\mathbf{k}_2,0) \right]
  \right\rangle, \nonumber
\end{align}
with
\begin{align}
  M^{ab}_{\alpha\beta\gamma\delta}&=\frac{1}{2}\left(\hat{n}^a\times\left(\hat{n}^a\times\mathbf{\sigma}_{\alpha\beta}\right)\right)\cdot\frac{1}{2}\left(\hat{n}^b\times\left(\hat{n}^b\times\mathbf{\sigma}_{\gamma\delta}\right)\right),
\end{align}
Lastly, we transform our basis to the band basis using the results
from our Hartree-Fock mean-field calculation, evaluate the two-body
expectation value via Wick's theorem, and Fourier transform to
frequency space to obtain
$\mathcal{\chi}^{ab}_{\text{MF}\perp}(\mathbf{q},\omega)$.



\bibliography{pyrochlore_spinwaves}

\end{document}